\documentclass{article}

\usepackage{amsmath, amsfonts, amssymb, amsxtra, amsopn}
\usepackage{pgfplots}
\pgfplotsset{compat=1.13}
\usepackage{pgfplotstable}
\usepackage{graphicx}
\usepackage{multirow}
\usepackage{algorithm}
\usepackage{algorithmicx}
\usepackage{algpseudocode}
\usepackage{booktabs}
\usepackage{listings}
\usepackage{cmap}
\usepackage{colortbl}
\usepackage{adjustbox}
\usepackage{epsfig}

\usepackage[tableposition=top]{caption}

\PassOptionsToPackage{hyphens}{url}

\usepackage{hyperref}
\hypersetup{colorlinks=true,linkcolor=black,citecolor=black,urlcolor=blue,filecolor=black}
\hypersetup{pdfpagemode=UseNone,pdfstartview=}

\pgfkeys{
    /pgf/number format/precision=4, 
    /pgf/number format/fixed zerofill=true }
    
\pgfplotstableset{
    /color cells/min/.initial=0,
    /color cells/max/.initial=1000,
    /color cells/textcolor/.initial=,
    color cells/.code={%
        \pgfqkeys{/color cells}{#1}%
        \pgfkeysalso{%
            postproc cell content/.code={%
                \begingroup
                \pgfkeysgetvalue{/pgfplots/table/@preprocessed cell content}\value
\ifx\value\empty
\endgroup
\else
                \pgfmathfloatparsenumber{\value}%
                \pgfmathfloattofixed{\pgfmathresult}%
                \let\value=\pgfmathresult
                \pgfplotscolormapaccess
                    [\pgfkeysvalueof{/color cells/min}:\pgfkeysvalueof{/color cells/max}]%
                    {\value}%
                    {\pgfkeysvalueof{/pgfplots/colormap name}}%
                \pgfkeysgetvalue{/pgfplots/table/@cell content}\typesetvalue
                \pgfkeysgetvalue{/color cells/textcolor}\textcolorvalue
                \toks0=\expandafter{\typesetvalue}%
                \xdef\temp{%
                    \noexpand\pgfkeysalso{%
                        @cell content={%
                            \noexpand\cellcolor[rgb]{\pgfmathresult}%
                            \noexpand\definecolor{mapped color}{rgb}{\pgfmathresult}%
                            \ifx\textcolorvalue\empty
                            \else
                                \noexpand\color{\textcolorvalue}%
                            \fi
                            \the\toks0 %
                        }%
                    }%
                }%
                \endgroup
                \temp
\fi
            }%
        }%
    }
}

\usepackage{enumitem}
\setlist[itemize]{noitemsep, topsep=0pt}

\advance\oddsidemargin by -0.2in
\advance\textwidth by 0.4in

\advance\topmargin by -0.25in
\advance\textheight by 0.5in

\long\def\symbolfootnotetext[#1]#2{\begingroup%
\def\thefootnote{\fnsymbol{footnote}}\footnotetext[#1]{#2}\endgroup}

\def\zz{\phantom{0}}

\algrenewcommand{\algorithmiccomment}[1]{{\color{red}{\tt //}\ #1}}

\algnewcommand{\Initialize}[1]{%
  \State \textbf{Initialize:}
  \Statex \hspace*{\algorithmicindent}\parbox[t]{.8\linewidth}{\raggedright #1}
}
\algnewcommand{\Given}[1]{%
  \State \textbf{Given:}
  \Statex \hspace*{\algorithmicindent}\parbox[t]{.8\linewidth}{\raggedright #1}
}

\def\sscoin{%
  \leavevmode
  \vtop{\offinterlineskip 
    \setbox0=\hbox{\scriptsize S}%
    \setbox2=\hbox to\wd0{\hfil\hskip-.03em
    \vrule height .3ex width .08ex\hskip .08em
    \vrule height .3ex width .08ex\hfil}
    \vbox{\copy2\box0}\box2}}
\newcommand\affil[2]{%
  \begingroup
  \renewcommand\thefootnote{}\footnote{\llap{$\hbox{}^{#1}\hbox{}$}#2}%
  \addtocounter{footnote}{-1}%
  \endgroup
}
\newcommand\markonly[1]{%
$\hbox{}^{\mbox{\kern4.5pt,\kern0.75pt #1}}$
}

\advance\oddsidemargin by -0.525in
\advance\textwidth by 1.05in
\advance\topmargin by -0.625in
\advance\textheight by 1.25in

\title{\vspace{-0.5in}Universal Adversarial Perturbations and Image Spam Classifiers}

\author{Andy Phung\footnote{phungandy0080@students.esuhsd.org}\ \ \ \ 
Mark Stamp\footnote{mark.stamp@sjsu.edu}\markonly{\sscoin}
}

\date{}

\begin{document}

\maketitle

\vglue-0.35in

\affil{\sscoin}{Department 
of Computer Science,
San Jose State University,
San Jose, California}

\abstract{As the name suggests, image spam is spam email 
that has been embedded in an image. Image spam was developed
in an effort to evade text-based filters. Modern deep learning-based classifiers
perform well in detecting typical image spam that is seen in the wild. 
In this chapter, we evaluate numerous adversarial techniques 
for the purpose of attacking deep learning-based image spam classifiers.
Of the techniques tested, we find that universal perturbation performs best. 
Using universal adversarial perturbations, we propose and analyze a new transformation-based 
adversarial attack that enables us to create tailored ``natural perturbations'' in image spam.
The resulting spam images benefit from both the presence of concentrated natural features and 
a universal adversarial perturbation. We show that the proposed technique outperforms 
existing adversarial attacks in terms of accuracy reduction, computation time per example, 
and perturbation distance. We apply our technique to create a dataset of adversarial 
spam images, which can serve as a challenge dataset for future research in 
image spam detection.}

\section{Introduction}

E-mail, or electronic mail, is one of the most popular forms of communication in the world, with
over~3.9 billion active email users~\cite{campaign_monitor}. As a side effect of this rapid growth, 
the number of unwanted bulk email messages---i.e., spam messages---sent with commercial 
or malicious intent has also grown. According to~\cite{campaign_monitor},
60 billion spam emails will be sent each 
day for the next three years. 

While text-based spam filtering systems are in use by most, 
if not all, e-mail clients~\cite{machine_learning_for_email_spam_filtering}, spammers 
can embed messages in attached images to evade such systems---such messages are known as 
image spam. Image spam detectors based on optical character recognition (OCR) have been 
deployed to combat such e-mail. As a countermeasure, spammers can modify images 
so as to disrupt OCR based techniques~\cite{image_spam_hunter}. 

In recent years, deep learning models, such as multi-layer perceptrons and convolutional neural networks, 
have been successfully applied to the image spam problem~\cite{image_spam_hunter, 
image_spam_analysis_and_detection, statistical_feature_extraction_for_classification, 
support_vector_machines_for_image_spam, deepimagespam, image_spam_filtering_using_conv, 
convolutional_neural_networks_for_image_spam}. Note that these techniques do not rely
on OCR, but instead detect image spam directly, based on characteristics of the images.

With the recent development of perturbation methods, the possibility exists for spammers 
to utilize adversarial techniques to defeat image-based machine learning 
detectors~\cite{intriguing_properties}. To date, we are not aware of perturbation techniques
having been used by image spammers, but it is highly likely that this will occur in the
near future.

The main contributions of our research are the following.
\begin{itemize}
\item We show that the universal perturbation adversarial attack is best suited for the 
task of bypassing deep learning-based image spam filters.
\item We propose a new image transformation-based attack that utilizes the maximization 
of layer activations to produce spam images containing universal perturbations. This technique
focuses perturbations in the most salient regions, as well as concentrating natural features 
in the remaining regions.
\item We compare our proposed adversarial technique to existing attacks and find that our 
approach outperforms all others in terms of accuracy reduction, computation time per example, 
and perturbation magnitude.
\item We generate a large dataset containing both non-spam and adversarial spam images using 
our proposed attack. The authors will make this dataset available to researchers.
\end{itemize}

The remainder of this chapter is organized as follows. In Section~\ref{sec:bac}, we provide
an overview of relevant research and related work.
In Section~\ref{sec:adv}, we evaluate adversarial attacks in the context of image spam, 
and in Section~\ref{sec:univ}, we present our proposed attack.
Finally, Section~\ref{sec:con} concludes this chapter, where we have included
suggestions for future work.


\section{Background}\label{sec:bac}

\subsection{Image Spam Filtering}

The initial defenses against image spam relied on
optical character recognition (OCR). In such OCR-based systems, 
text is extracted from an image, at which point
a traditional text-based spam filter can be used~\cite{spam_assassin}. 
As a reaction to OCR-based techniques,
spammers introduced images with slight modifications, such as 
overlaying a light background of random artifacts on images, which are
sufficient to render OCR ineffective. The rise of learning algorithms, however, 
has enabled the creation of image spame filtering systems based directly on 
image features. 

In~2008, a filtering system using a global image feature-based probabilistic boosting 
tree was proposed, and achieved an~89.44\%\ detection rate with a false positive rate 
of~0.86\%~\cite{image_spam_hunter}. Two years later, an artificial neural network for 
image classification was proposed~\cite{statistical_feature_extraction_for_classification}. 
These latter authors used were able to classify image spam
with~92.82\%\ accuracy based on color histograms, 
and~89.39\%\ accuracy based on image composition extraction. 
 
The two image spam detection methods presented 
in~\cite{image_spam_analysis_and_detection} rely on principal component analysis (PCA)
and support vector machines (SVM). In addition, the authors 
of~\cite{image_spam_analysis_and_detection} introduce a new dataset that their 
methods cannot reliably detect. 
Two years later, the authors of~\cite{support_vector_machines_for_image_spam} improved 
on the results in~\cite{image_spam_analysis_and_detection} by training a linear SVM 
on~38 image features, achieving~98\%, 
accuracy in the best case. The authors also introduce a challenge dataset that is even 
more challenging than the analogous dataset presented 
in~\cite{image_spam_analysis_and_detection}. 
 
The recent rise of deep learning, a subfield of machine learning, coupled with advances 
in computational speed has enabled the creation of filtering systems capable of considering 
not only image features, but entire images at once. In particular, 
convolutional neural networks (CNN) 
are well suited to computer vision tasks due to their powerful feature extraction capabilities. 

In recent years, CNNs have been applied to the task of image spam detection.
For example, in~\cite{image_spam_filtering_using_conv} a CNN is trained on 
an augmented dataset of spam images, achieving a~6\%\ improvement in accuracy,
as compared to previous work. Similarly, the authors of~\cite{deepimagespam} 
consider a CNN, which achieved~91.7\%\ accuracy. 
In~\cite{convolutional_neural_networks_for_image_spam},
a CNN-based system is proposed, which achieves an accuracy of~99\%\ on 
a real-world image spam dataset,~83\%\ accuracy  
on the challenge dataset in~\cite{image_spam_analysis_and_detection}
(an improvement over previous works), 
and~68\%\ on the challenge dataset in~\cite{support_vector_machines_for_image_spam}. 

From the challenge datasets introduced in~\cite{image_spam_analysis_and_detection} and~\cite{support_vector_machines_for_image_spam}, we see that
the accuracy of machine learning-based filtering systems can be 
reduced significantly with appropriate modifications to spam images. In this research, 
we show that the accuracy of 
such systems can be reduced far more by using the adversarial learning
approach that we present below. 

\subsection{Adversarial Learning}

The authors of~\cite{intriguing_properties} found that by applying an imperceptible 
filter to an image, a given neural network's prediction can be arbitrarily changed. 
This filter can be generated from the optimization problem
\begin{equation}\nonumber
    \begin{split}
&    \textbf{minimize } \|r\|_2 \\
&\textbf{subject to } f(x+r) = l \text{ and } x+r \in [0, 1]^m
    \end{split}
\end{equation}
where $f$ is the classifier, $r$ is the minimizer, $l$ is the target label, and $m$ is the dimension of the image.  
The resulting modified images are said to be \textit{adversarial examples}, and the attack presented in~\cite{intriguing_properties} is known as the \textit{L-BFGS Attack}. These adversarial examples generalize well to different network architectures and networks.
 
More recently, many advances have been made in both adversarial example generation and 
detection. For example, in~\cite{adversarial_examples_attacks_and_defenses} a taxonomy 
is proposed
for generation and detection methods, as well as a threat model. Based on this threat model, 
the task of attacking neural network-based image spam detectors requires an attack that is 
false-negative (i.e. generative of positive samples misclassified as negative) and black-box 
(i.e. the attacker does not have access to the trained model). 
Attacks on image spam classifiers must satisfy these two criteria. 

After the introduction of the L-BFGS Attack, the authors of~\cite{explaining_and_harnessing} 
built on their work in~\cite{intriguing_properties} by introducing 
the \textit{Fast Gradient Sign Method} (FGSM). This method uses the gradient of the 
loss function  with respect to a given input image to efficiently create a new image that 
maximizes the loss, via backpropagation. This can be summarized with the expression
$$
    \mbox{adv}_x = x + \epsilon\,\mbox{sign}\bigl(\nabla_x J(\theta, x, y)\bigr)
$$
where $\theta$ is the parameters of the model, $x$ is the input image, $y$ is the target label, 
and $J$ is the cost function used to train the model.
These authors also introduce the notion that adversarial examples result from linear 
behavior in high-dimensional spaces. 

The authors of~\cite{towards_evaluating_the_robustness} introduce \textit{C\&W's Attack}, 
a method designed to combat \textit{defensive distillation}, which consists of training a 
pair of models such that there is a low probability of successively attacking both models. 
C\&W's Attack is a non-box constrained variant of the L-BFGS Attack that is more easily 
optimized and effective against both distilled and undistilled networks. 
They formulate adversarial example generation as the optimization problem
\begin{equation}\nonumber
  \begin{split}
& \textbf{minimize } D(x, x + \delta) + c \cdot f(x + \delta) \\ 
& \textbf{such that } x + \delta \in [0, 1]^n  
\end{split}  
\end{equation}
where $x$ is the image, $D$ is one of the three distance metrics described below, and $c$ is a suitably chosen constraint (the authors choose $c$ with binary search).  
The authors also utilize three distance metrics for measuring perturbation: $L_0$ (the number of altered pixels), $L_2$ (the Euclidean distance), and $L_{\infty}$ (the maximum change to any of the coordinates), and introduced three subvariants of their attack that aim to minimize each of these distance metrics. 
 
It is important to note that the previously mentioned attacks require knowledge of the classifier's gradient and, as such, cannot be directly deployed in a black-box attack. In~\cite{practical_black_box}, the authors propose using a surrogate model for adversarial example generation to enable the transferability of adversarial examples to attack black-box models. Differing from gradient-based methods, the authors of~\cite{zoo} introduced a method, 
\textit{Zeroth Order Optimization} (ZOO), which is inspired by
the work in~\cite{towards_evaluating_the_robustness}.
The ZOO technique employs gradient estimation, with the most significant downside 
being that it is computationally expensive. 

The paper~\cite{deepfool} introduces the \textit{DeepFool} attack, which aims to find the minimum
distance from the original input images to the decision boundary for adversarial examples. 
They found that the minimal perturbation needed for an affine classifier is the distance to the separating affine hyperplane, which is expressed (for differentiable binary classifiers) as 
\begin{equation}\nonumber
\begin{split}
& \textbf{argmin}_{\eta_i} \|\eta_i\|_2 \\
& \textbf{such that } f(x_i) + \nabla f(x_i)^T \eta_i = 0  
\end{split}
\end{equation}
where $i$ denotes the iteration, $\eta$ is the perturbation, and $f$ is the classifier.
In comparison to FGSM, DeepFool minimizes the magnitude of the perturbation, instead of the number  of selected features. This would appear to be ideal for spammers, since it would tend to minimize the effect on an image.
 
The \textit{universal perturbation} attack presented in~\cite{universal} is also suited to 
the task at hand. We believe that universal adversarial examples are most likely 
to be deployed by spammers against black-box models due to their simplicity 
and their transferability across architectures. Generating universal 
perturbations is an iterative process, as the goal is to find a vector $v$ that satisfies 
$$
    \|v\|_p \leq \xi \mbox{\ \ and\ \ }
    \mathbb{P}_{x\sim \mu} (\hat{k}(x+v) \neq \hat{k}(x)) \geq 1 - \delta
$$
where $\mu$ is a distribution of images, $\hat{k}$ is a 
classification function that outputs for each image $x$ and a label $\hat{k}(x)$.
The results in~\cite{universal} show that universal perturbations are misclassified with high probability, suggesting that the existence of such perturbations are correlated to certain regions of the decision boundary of a deep neural network.

Finally, the authors of~\cite{restoration_as_a_defense} propose input restoration with a 
preprocessing network to defend against adversarial attacks. 
The authors' defense improved the classification precision of a CNN 
from~10.2\%\ to~81.8\%, on average. These results outperform 
existing input transformation-based defenses. 

\section{Evaluating Adversarial Attacks}\label{sec:adv}

\subsection{Experimental Design}

The two multi-layer perceptron and convolutional neural network architectures presented in~\cite{convolutional_neural_networks_for_image_spam} are each trained on both of
the dataset presented in~\cite{image_spam_hunter}, which henceforth will be referred to 
as the \textit{ISH Dataset}, and the dataset presented 
in~\cite{support_vector_machines_for_image_spam}, which henceforth will be referred to as 
the \textit{MD Dataset} (modified Dredze). We use TensorFlow~\cite{tensorflow} to train
our models---both architectures have been trained as they were presented in their 
respective articles on each of the datasets. NumPy~\cite{numpy} and 
OpenCV~\cite{opencv} are used for numerical operations and image processing 
tasks, respectively. All computation are performed on a
laptop with~8GB ram, using Google Colaboratory's Pro GPU. 

The ISH Dataset contains~928 spam images and~830 non-spam images, 
while the MD Dataset contains~810 spam images and~784 non-spam images;
all images in both datasets are in \textit{jpg} format. These datasets are summarized
in Table~\ref{tab:data}.

\begin{table}[!htb]
\caption{Image spam datasets}\label{tab:data}
\centering
\begin{tabular}{ccc}
\midrule\midrule
\textbf{Name} & \textbf{Spam images} & \textbf{Non-spam images} \\
\midrule
ISH dataset & \zz928 & \zz830 \\
MD dataset & \zz810 & \zz784 \\
\midrule
Total & 1738 & 1613 \\
\midrule\midrule
\end{tabular}
\end{table}

Dataset preprocessing for the networks 
presented in~\cite{convolutional_neural_networks_for_image_spam} consist 
of downsizing each of the images such that their dimensions are~$32\time 32\times 3$, 
applying zero-parameter Canny edge detection~\cite{canny} to a copy of the downsized 
image, and concatenating the downsized image with the copy that had 
Canny edge detection applied. This process results in~$64\time 32\time 3$ images, 
which are used to train the two neural networks, one for the ISH dataset, and one for the MD dataset. 
The four resulting models achieved accuracies within roughly~7\%\ of the accuracies 
reported in~\cite{convolutional_neural_networks_for_image_spam}. 

To enable the generation of adversarial examples, four larger models with an input size of 400x400 are also trained on 
the original datasets. The first few layers of each of these models are simply used to 
downscale input images such that the original architectures can be used after downscaling. 
These four alternative models achieve accuracy roughly equivalent to the 
original models. The four adversarial attacks 
(FGSM, C\&W’s Attack, DeepFool, and Universal Perturbation) 
utilize these four alternative models to generate adversarial examples that can then be formatted 
as the original datasets to attack the original four models. This procedure attempts to exploit 
the transferability of adversarial examples to similar architectures. 

The IBM Adversarial Robustness Toolbox (ART)~\cite{adversarial_robustness_toolbox} is 
used to implement C\&W’s Attack, DeepFool, 
and Universal Perturbations, while FGSM was implemented independently from scratch.
An attempt was made to optimize the parameters of each technique---the resulting
parameters are summarized in Table~\ref{tab:parms}. Note that
for the Universal Perturbation attack, FGSM was used as the base attack, as the IBM ART 
allows any adversarial attack to be used for computing universal perturbations.

\begin{table}[!htb]
\advance\tabcolsep by 4pt
\caption{Attack parameters}\label{tab:parms}
\centering
\begin{tabular}{c|ll}
\midrule\midrule
\textbf{Attack} & \textbf{Description} & \textbf{Value} \\
\midrule
FGSM 
  & perturbation magnitude & 0.1 \\ \midrule
\multirow{6}{*}{C\&W's attack} 
  & target confidence & 0 \\ 
  & learning rate & 0.001 \\
  & binary search steps & 20 \\ 
  & maximum iterations & 250 \\
  & initial trade-off & 100 \\
  & batch size & 1 \\ \midrule
\multirow{4}{*}{DeepFool}
  & max iterations & 500 \\ 
  & overshoot parameter & $10^{-6}$ \\ 
  & class gradients & 10 \\
  & batch size & 1 \\ \midrule
\multirow{4}{*}{Universal Perturbation}
  & target accuracy & 0\% \\ 
  & max iterations & 250 \\
  & step size & 64 \\
  & norm & $\infty$ \\
\midrule\midrule
\end{tabular}
\end{table}

The metrics used to evaluate each of the four attacks are the average accuracy, 
area under the curve (AUC) of the receiver operating characteristic (ROC) curve, average~$L_2$ perturbation measurement 
(Euclidean distance), and average computation time per example for each of the four models. 
Scikit-learn~\cite{scikit-learn} was used to generate the ROC curves for each attack. 

We use~251 data points for accuracy 
and $L_2$ distances collected for the FGSM, DeepFool, and Universal Perturbation experiments, 
in accordance with the full size of the test dataset, which contains~251 examples for generating adversarial examples. However, only~$28$ data points were 
collected from the C\&W's Attack experiment due to 
the large amount of time required to generate each data point (roughly five minutes per data point). 
The technique that will be used as the basis of our proposed attack will be selected based 
on the performance of each attack, as presented in the next section.

\subsection{Analysis}


The mean accuracy, computation time per example, and~$L_2$ distance were 
recorded for each of the four models attacked by each of the attack methods. 
This data was compiled into the tables discussed in this section. 

\begin{figure}[!htb]
\centering
\includegraphics[width=7.8cm]{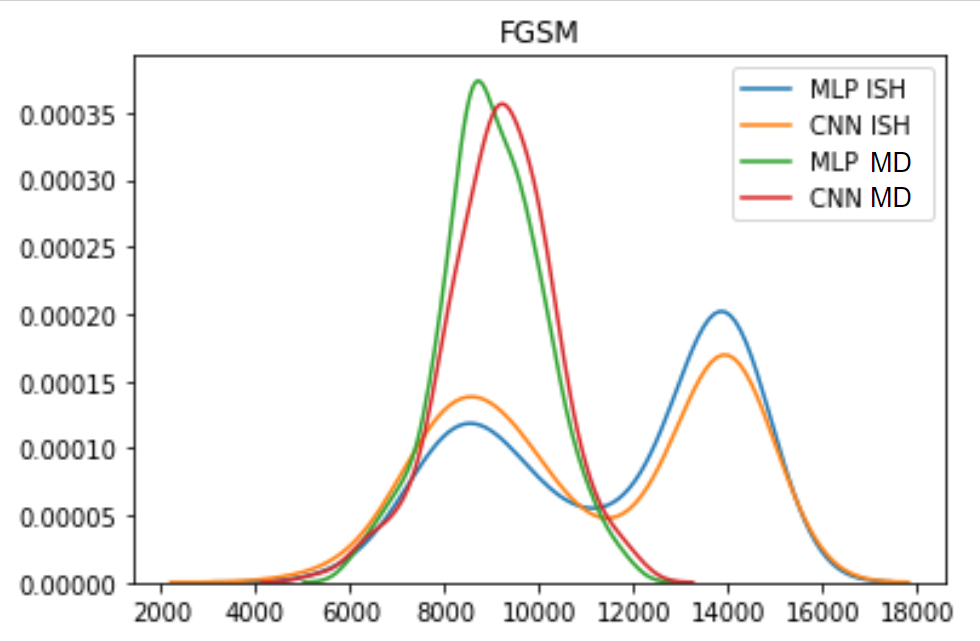}
\caption{Density plot of~$L_2$ (Euclidean) distances (Fast Gradient Sign Method)}
\label{fig:fgsml2}
\centering
\end{figure}

From Table~\ref{tab:meanacc1} we see that for FGSM, the accuracy of the attacked models 
is shown to vary inconsistently while Figure~\ref{fig:fgsml2} shows that the distribution of 
the~$L_2$ distances of the generated adversarial examples skew right. 
Based on these results 
and corresponding density plots of the accuracy and~$L_2$ distance distributions, 
the FGSM attack can be ruled out as a candidate due to poor accuracy. 

\begin{table}[!htb]
\advance\tabcolsep by 4pt
    \centering
    \caption{Mean accuracy per adversarial example}
    \label{tab:meanacc1}
    \begin{tabular}{c|cccc}
     \midrule\midrule
      \multirow{2}{*}{\textbf{Model}}
       & \multirow{2}{*}{\textbf{FGSM}}
       & \textbf{C\&W's}
       & \multirow{2}{*}{\textbf{DeepFool}}
       & \textbf{Universal}\\
                             &                          &  \textbf{Attack} &                              &  \textbf{Perturbation}\\
     \midrule
     MLP (ISH) & 95.2\% & 89.2\% & 98.8\% & 98.7\%\\
     CNN (ISH) & 36.2\% & 49.6\% & 61.5\% & 49.9\%\\
     MLP (MD) & 69.7\% & 75.6\% & 93.5\% & 94.3\%\\
     CNN (MD) & 82.8\% & 77.2\% & 14.5\% & \zz8.4\%\\
     \midrule\midrule
    \end{tabular}
\end{table}

The mean~$L_2$ (Euclidean) distances of the adversarial examples
are given in Table~\ref{tab:meandist1}. The distribution of distances appears to be roughly equivalent across all attacks.

\begin{table}[!htb]
\advance\tabcolsep by 4pt
    \centering  
    \caption{Mean $L_2$ (Euclidean) distance of adversarial examples from original images}
    \label{tab:meandist1}
    \begin{tabular}{c|cccc}
     \midrule\midrule
      \multirow{2}{*}{\textbf{Model}}
       & \multirow{2}{*}{\textbf{FGSM}}
       & \textbf{C\&W's}
       & \multirow{2}{*}{\textbf{DeepFool}}
       & \textbf{Universal}\\
                             &                          &  \textbf{Attack} &                              &  \textbf{Perturbation}\\
     \midrule
     MLP (ISH) & 11537.55 & 10321.77 & 11513.26 & 11483.72\\
     CNN (ISH) & 11108.44 & 10924.14 & 11216.19 & 11416.58\\
     MLP (MD) & \zz8998.71 & \zz9185.04 & \zz9566.02 & \zz9490.56\\
     CNN (MD) & \zz9144.49 & \zz9009.91 & \zz9128.99 & \zz9381.15\\
     \hline
    \end{tabular}
\end{table}

DeepFool can also be ruled as a candidate, as the attack has been seen to be 
only marginally better than the FGSM attack in terms of performance, while also 
having a significantly higher average computation time per adversarial example. This
can be observed in Table~\ref{tab:meancomp1}, where the computation time 
per example varies greatly. 

\begin{table}[!htb]
\advance\tabcolsep by 4pt
    \centering
    \caption{Mean computation time per adversarial example}
    \label{tab:meancomp1}
    \begin{tabular}{c|cccc}
     \midrule\midrule
      \multirow{2}{*}{\textbf{Model}}
       & \multirow{2}{*}{\textbf{FGSM}}
       & \textbf{C\&W's}
       & \multirow{2}{*}{\textbf{DeepFool}}
       & \textbf{Universal}\\
                             &                          &  \textbf{Attack} &                              &  \textbf{Perturbation}\\
     \midrule
     MLP (ISH) & 0.180 & 269.65 & 19.90 & 4.37\\
     CNN (ISH) & 0.038 & 251.01 & \zz4.75 & 2.87\\
     MLP (MD) & 0.164 & 270.58 & 36.30 & 3.71\\
     CNN (MD) & 0.165 & 244.47 & \zz1.48 & 5.23\\
     \hline
    \end{tabular}
\end{table}

\begin{figure}[!htb]
\centering
\includegraphics[width=8cm]{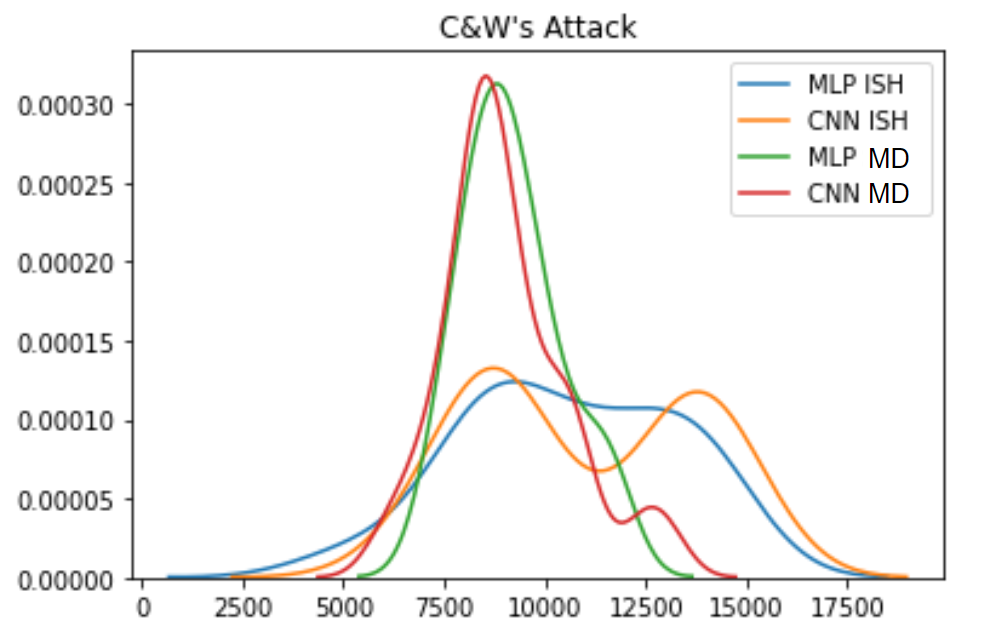}
\caption{Density plot of~$L_2$ (Euclidean) distances (C\&W's Attack)}
\label{fig:cwl2}
\centering
\end{figure}

In contrast, C\&W's Attack shows consistent performance in all three metrics at the cost of 
high computation time (roughly five minutes per adversarial example).
The consistency of this attack is ideal from a spammer's perspective, 
though the trade-off is a relatively high computation time. In addition, the 
left skew of this attack with respect to~$L_2$ distance, as presented in Figure~\ref{fig:cwl2}, 
indicates that the perturbation made to spam images is much lower in comparison to the other attacks.

The Universal Perturbation attack is inconsistent in terms of accuracy, as shown 
in Table~\ref{tab:meanacc1} where the mean accuracy across the four models is clearly 
shown to fluctuate wildly, but this 
is simply due to the fact that only one perturbation (albeit with varying success across architectures) is applied to all spam images, which is highly advantageous for spammers.
The generation and application of this perturbation to an image takes roughly four seconds,
which would result in greater performance in a real-world spam setting in comparison 
to C\&W's Attack. 
 
\begin{figure}[!htb]
  \centering
  \begin{minipage}[b]{0.45\textwidth}
    \includegraphics[width=\textwidth]{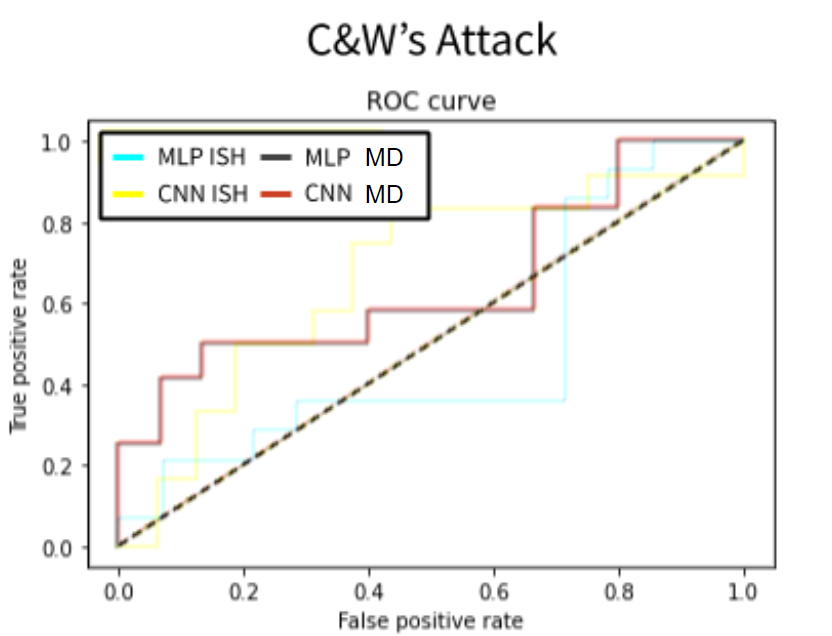}
  \end{minipage}
  \begin{minipage}[b]{0.45\textwidth}
    \includegraphics[width=\textwidth]{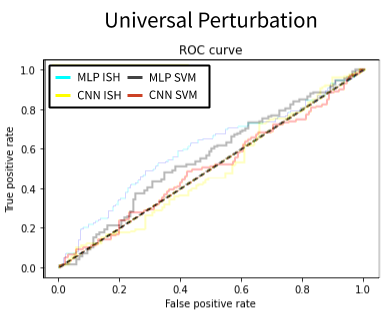}
  \end{minipage}
  \caption{ROC curves of C\&W's Attack and the Universal Perturbation when used to attack the four classifiers}
  \label{fig:cwroc_uproc}
\end{figure}

To further compare C\&W's Attack and the Universal Perturbation attack, 
the ROC curves of the two are presented in Figure~\ref{fig:cwroc_uproc}.
These ROC curves can be used to quantify the diagnostic ability of the 
models attacked by each method.

The ROC curve for C\&W's attack is much noisier due to being generated from only~28 
data points. Taking this into consideration, it can be inferred that both C\&W's Attack 
and the Universal Perturbation attack are able to reduce the areas under the ROC curve (AUC) 
of the attacked models to values close to~0.5. This suggests that both attacks are able to reduce the class separation capacity of attacked image spam classifiers to essentially random. 

To analyze the differences in distribution of the accuracy and~$L_2$ distance data collected from 
the trials conducted on C\&W's Attack and the Universal Perturbation attack, the Mann-Whitney U Test was utilized via its implementation 
in SciPy~\cite{scipy}. The Mann-Whitney U Test compares two populations---in 
this case, the accuracy and~$L_2$ distance data from both attacks for each attacked model.
The null hypothesis (H0) for the test is that the probability is~50\%\ that a randomly drawn 
value from the first population will exceed a value from the second population. The result of each 
test is a Mann-Whitney U Statistic (not relevant in our case) and a~$p$-value. We use
the~$p$-value to determine whether the difference between the data is statistically 
significant, where the standard threshold is~$p = 0.05$. The results of these tests are 
given in Table~\ref{tab:mannwhitneyu}.

\begin{table}[!htb]
\advance\tabcolsep by 4pt
    \centering
    \caption{Mann-Whitney U Test results comparing C\&W's Attack and the Universal Perturbation attack}
    \label{tab:mannwhitneyu}
    \begin{tabular}{l|ll}
     \midrule\midrule
     \multicolumn{1}{c|}{\textbf{Model}}
                 & \multicolumn{1}{c}{\textbf{Accuracy $p$-value}} 
                 & \multicolumn{1}{c}{\textbf{$L_2$ distance $p$-value}} \\
     \midrule
     MLP ISH & 0.000 (H0 is rejected) & 0.034 (H0 is rejected)\\
     CNN ISH & 0.384 (H0 is not rejected) & 0.098 (H0 is not rejected)\\
     MLP MD & 0.000 (H0 is rejected) & 0.057 (H0 is not rejected)\\
     CNN MD & 0.000 (H0 is rejected) & 0.016 (H0 is rejected)\\
     \midrule\midrule
    \end{tabular}
\end{table}

The results  in Table~\ref{tab:mannwhitneyu} imply that the performance of these two attacks (C\&W's Attack and the Universal Perturbation attack)
are nearly identical when attacking a CNN trained on the ISH dataset, 
as evidenced in the second row, where the null hypothesis is not rejected. 
However, the~$L_2$ distance measurement for spam images 
that have had the universal perturbation applied should remain constant relative 
to the original spam image. Therefore, the results of these tests suggest that 
the Universal Perturbation attack is able to achieve similar performance 
to C\&W's Attack, in terms of perturbation magnitude, 
with a much lower computation time per example in comparison to C\&W's Attack. 

\begin{figure}[!htb]
  \centering
    \includegraphics[width=8cm]{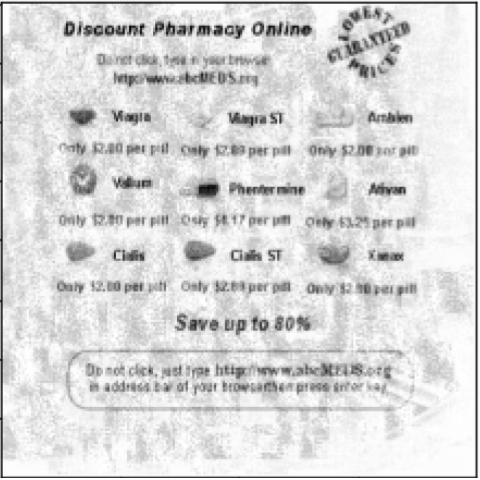}
    \caption{Adversarial spam image generated with the Universal Perturbation attack}
    \label{fig:sampleadv}
\end{figure}

Given the above evidence, the Universal Perturbation attack is the best choice for 
image spam, as it is unrivaled in terms of potential performance in a real-world 
setting. The key advantages of the Universal Perturbation attack include
that it generates a single perturbation to be applied to all spam images, 
and its relatively fast computation time per adversarial example. Therefore,
Universal Perturbation will be used as a basis for our image transformation 
technique, as discussed and analyzed in the remainder of this paper. A sample adversarial spam image generated with the Universal Perturbation attack is presented in Figure~\ref{fig:sampleadv}.


\section{Inceptionism-Augmented Universal Perturbations}\label{sec:univ}

\subsection{Procedure}

Based on the results and discussion above, a transformation that
is applied to spam images prior to generating adversarial examples, since perturbations cannot be transformed after application, should 
meet the following conditions.
\begin{itemize}
    \item Lower the misclassification rate
    \item Preserve adversarial effects after image resizing
    \item Make non-spam features more prominent, while retaining legibility
\end{itemize}
Given the above criteria, a reasonable approach would be to maximize the presence 
of "natural features" in a given spam image. That is, the features characteristic of non-spam images learned by classifiers should be maximized whilst retaining legibility. To accomplish this, the procedure for maximizing 
the activation of a given output neuron (in this case, the non-spam output neuron), 
as introduced in~\cite{deepdream}, dubbed "DeepDream", can be used to increase the number of natural features in all images from the non-spam subsets of the ISH and MD datasets.
This is accomplished by maximizing the activations of the convolutional layer 
with the greatest number of parameters and output layer in the corresponding CNNs. 
The resulting two sets of images that have had DeepDream applied (``dreamified'' images) are then grouped into batches of four images. The weighted average of the four images in each batch can then be taken to produce two 
processed non-spam datasets of images with high concentrations of natural features, as batches of greater than four images may result in high noise. Each of the images in the resulting two non-spam datasets 
are henceforth referred to as \textit{natural perturbations}.

To preserve the adversarial effect that the universal perturbation introduces, 
the Gradient-weighted Class Activation Mapping (Grad-CAM) technique 
introduced in~\cite{gradcam} 
is used to generate a class activation map for each spam 
image in each dataset. The inverse of each such map is used with a natural perturbation 
generated from the same dataset to remove the regions of the 
natural perturbation where the class activation map is highest. 
By superimposing the resulting natural perturbations onto the corresponding spam images, 
the regions where the universal perturbation is most effective are left intact while the 
regions of the spam images affected by the natural perturbations benefit by being
more non-spam-like. The presence of natural features in the resulting spam images 
should also result in robustness against resizing prior to inference by a deep 
learning-based image spam detection model, as the natural features should be still be somewhat preserved even after being shrunken. 

The universal perturbation is then applied to each of the resulting spam images. 
The result is that we potentially reduce a deep learning-based image spam 
detector's accuracy due to the presence of a natural perturbation and a universal adversarial perturbation and retain some sort of adversarial effect in the case of resizing.
This procedure also allows for the retention of legible text within spam images. 

\subsection{Implementation}

To generate our two sets of ``dreamified'' images, the CNN architecture presented in~\cite{convolutional_neural_networks_for_image_spam} is trained on both 
the ISH and MD datasets, with inverted labels to allow for the maximization of the 
activations of the neurons corresponding to non-spam images, as the activations for spam images would be maximized if the labels weren't inverted. 
These two models are trained with the TensorFlow Keras API, with the 
hyperparameters given in~\cite{convolutional_neural_networks_for_image_spam}. 
For each of the models, the convolutional layer with the highest number of parameters 
and the output layer were chosen as the layers in which the activation should be maximized 
via gradient ascent, as the aforementioned convolutional layer is responsible for recognizing the most complex natural features. Each of the images from the non-spam subsets of the ISH and MD datasets were used for inference on the two CNN models. The CNN models use 
the losses of the chosen layers to iteratively update the non-spam images with gradient 
ascent so that the number of non-spam features is maximized. 
Each non-spam image 
is updated for~64 iterations with an update size of~$0.001$. The resulting ``dreamified'' 
images are then grouped into batches of~4 and blended via evenly distributed weighted 
addition to produce a total of~392 grayscale images, each of size~$400\times 400\times 1$.
These~392 grayscale images are evenly split between the ISH dataset and MD datasets.
 
To utilize GradCAM, 
the CNN architecture presented in~\cite{convolutional_neural_networks_for_image_spam} 
is trained on both the ISH and MD datasets with normal labels. 
For each image from the spam subsets of the ISH and MD datasets, 
GradCAM is used to generate a corresponding class activation map based 
on the activations of the last convolutional layer in each of the two models.
This is accomplished by computing the gradient of the top predicted class with 
respect to the output feature map of the last convolutional layer, 
using the mean intensity of the gradient over specific feature map channels. 
OpenCV~\cite{opencv} is then used to upscale each of the class activation maps 
to~$400\times 400$, convert them to binary format, and invert the result to allow the class activation maps to be applied to the natural perturbations such that only the areas with highest activation will contain the natural perturbations. 
The bitwise AND of each processed class activation map and a randomly selected 
natural perturbation can then be used to generate two sets of processed 
natural perturbations, which are superimposed on the corresponding spam images 
from each of the two spam subsets. This procedure results in two subsets 
of spam images with natural perturbations. 

Lastly, the universal perturbation is generated and applied to all images within 
the two spam image subsets that have had natural perturbations applied. For this
operation, we use the IBM Adversarial Robustness Toolbox~\cite{adversarial_robustness_toolbox}.
The hyperparameters for the Universal Perturbation attack remain the same 
as those given in Table~\ref{tab:parms}, above.

\subsection{Performance Evaluation}

The mean accuracy, computation time per example, and~$L_2$ distance were recorded for 
each of the four models attacked using spam images with modified universal perturbations.
This is analogous to what was done during the attack selection process. 
This data has been compiled into the tables discussed in this section.

As can be seen from the results in Table~\ref{tab:meanacc2},
the proposed method for generating adversarial 
spam images is capable of lowering a learning-based model's accuracy 
to~23.7\%. In addition, on average, our proposed technique is
much more effective while being evenly distributed in terms of accuracy 
on similar learning-based models. 

\begin{table}[!htb]
\advance\tabcolsep by 4pt
    \centering
    \caption{Mean accuracy of each model with spam images created by the proposed method}
    \label{tab:meanacc2}
    \resizebox{0.85\textwidth}{!}{
    \begin{tabular}{c|cccc}
     \midrule
     \midrule
      \textbf{Images} & \textbf{MLP (ISH)}  & \textbf{CNN (ISH)} & \textbf{MLP (MD)} & \textbf{CNN (MD)} \\
     \midrule
     Modified spam images & 80.1\% & 98.8\% & 98.4\% & 75.3\% \\
     Modified spam images with & \multirow{2}{*}{72.2\%} 
     						& \multirow{2}{*}{50.4\%} 
						& \multirow{2}{*}{78.7\%} 
						& \multirow{2}{*}{23.7\%} \\
     Universal Perturbations \\
     \midrule\midrule
    \end{tabular}}
\end{table}

From Table~\ref{tab:meancomp2}, we see that
in contrast to C\&W's Attack, which on average takes 258.93 seconds per example, the time necessary to generate adversarial spam images 
with natural perturbations is significantly lower and comparable to 
that of the original Universal Perturbation attack.
This is another advantage of our proposed attack.

\begin{table}[!htb]
\advance\tabcolsep by 4pt
    \centering
    \caption{Mean computation time per adversarial spam image (in seconds)}
    \label{tab:meancomp2}
    \begin{tabular}{cccc}
     \midrule\midrule
      \textbf{MLP (ISH)}  & \textbf{CNN (ISH)} & \textbf{MLP (MD)} & \textbf{CNN (MD)} \\
     \midrule
      5.46 & 5.15 & 5.87 & 4.80 \\
     \midrule\midrule
    \end{tabular}
\end{table}

\begin{figure}[!htb]
  \centering
    \includegraphics[width=8cm]{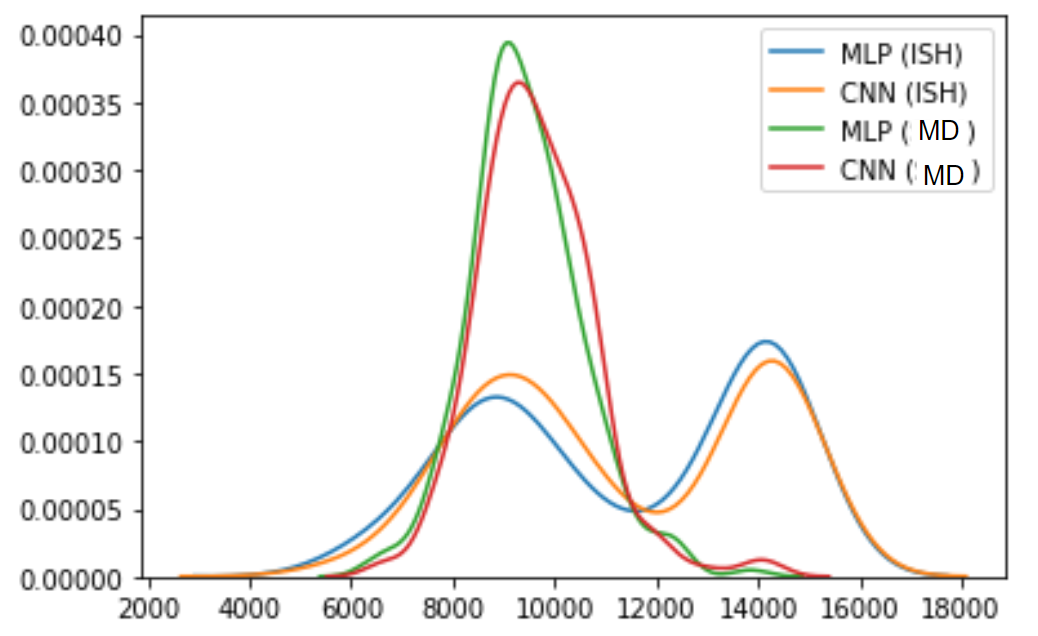}
    \caption{Density plot of~$L_2$ (Euclidean) distances of the modified adversarial spam images from the original images}
    \label{fig:lastl2}
\end{figure}

The mean~$L_2$ distances and the distribution of the~$L_2$ distances of the modified adversarial spam 
images are given in Table~\ref{tab:meandist2}. From Figure~\ref{fig:lastl2}, we see that the distributions 
of these distances are, on average, not skewed, indicating that the natural perturbations have 
had a slight negative effect on the spam image $L_2$ distances, as the distributions for the original Universal Perturbation attack were skewed to the left. 

\begin{table}[!htb]
\advance\tabcolsep by 4pt
    \centering
    \caption{Mean~$L_2$ (Euclidean) distance of modified adversarial spam images from original images}
    \label{tab:meandist2}
    \begin{tabular}{cccc}
     \midrule\midrule
      \textbf{MLP (ISH)}  & \textbf{CNN (ISH)} & \textbf{MLP (MD)} & \textbf{CNN (MD)} \\
     \midrule
      11392.02 & 11309.40 & 9440.69 & 9628.61 \\
     \midrule\midrule
    \end{tabular}
\end{table}

\begin{figure}[!htb]
  \centering
    \includegraphics[width=8cm]{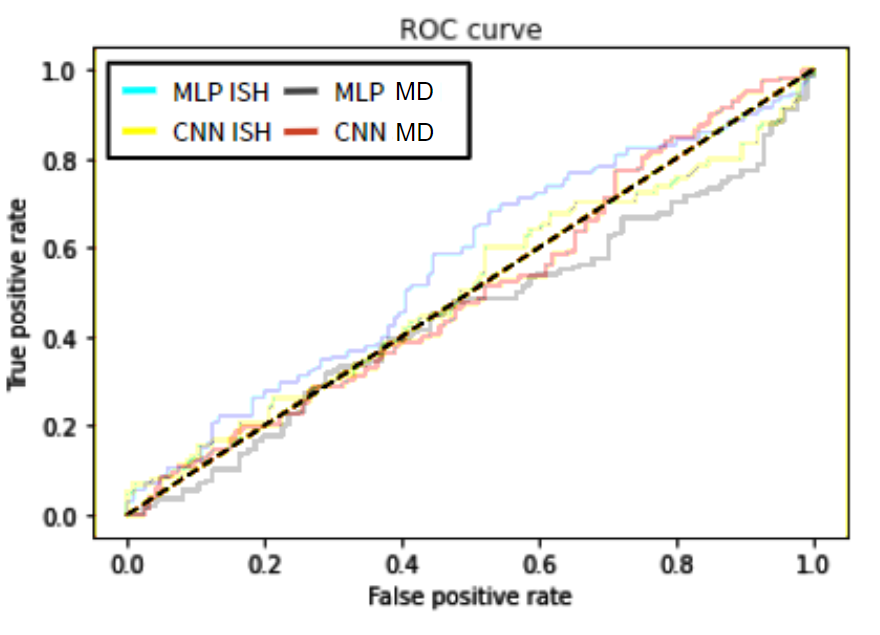}
    \caption{ROC curves of each of the four models attacked by the modified spam images generated with the proposed method}
    \label{fig:lastroc}
\end{figure}

The ROC curves of the models attacked by the proposed method, 
which appear in Figure~\ref{fig:lastroc}, are slightly worse in comparison to 
that of the original Universal Perturbation attack, suggesting once more that the 
attack is capable of reducing the class separation capacity of attacked image spam 
classifiers to essentially random.

\subsection{Proposed Dataset Analysis}

\begin{figure}[!htb]
  \centering
    \includegraphics[width=6cm]{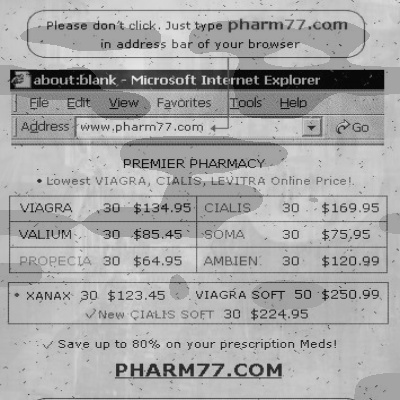}
    \caption{Example of modified adversarial spam image generated with the proposed method}
    \label{fig:spamimage2}
\end{figure}

Figure~\ref{fig:spamimage2} contains an example of a modified adversarial spam images.
From this image, we observe that
the proposed method was able to effectively utilize class activation maps 
generated with GradCAM to selectively apply a random natural perturbation 
to the spam image. As discussed in the previous section, this decreases 
classification accuracy even prior to the application of a universal perturbation. 

\begin{figure}[!htb]
  \centering
  \begin{minipage}[b]{0.425\textwidth}
    \includegraphics[width=\textwidth]{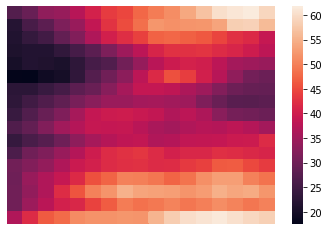}
    \caption{ISH spam data}
    \label{fig:ishheatmap1}
  \end{minipage}
  \begin{minipage}[b]{0.425\textwidth}
    \includegraphics[width=\textwidth]{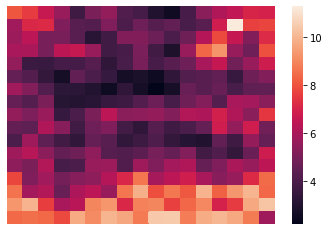}
    \caption{MD spam data}
    \label{fig:svmheatmap1}
  \end{minipage}
\end{figure}

\begin{figure}[!htb]
  \centering
  \begin{minipage}[b]{0.425\textwidth}
    \includegraphics[width=\textwidth]{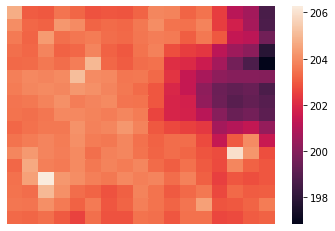}
    \caption{Modified ISH spam data}
    \label{fig:origishheatmap1}
  \end{minipage}
  \begin{minipage}[b]{0.425\textwidth}
    \includegraphics[width=\textwidth]{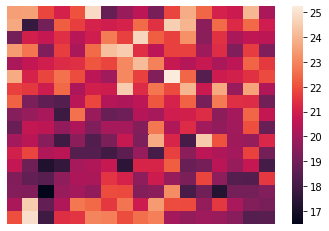}
    \caption{Modified MD spam data}
    \label{fig:origsvmheatmap1}
  \end{minipage}
\end{figure}

To fully evaluate the effect of the modified adversarial spam images from the two 
modified datasets, two sets of class activation maps are generated from the spam 
subsets of the two datasets using GradCAM and the corresponding CNN models.
These activation maps are then averaged to obtain two heatmaps 
from the class activation maps, as shown in Figures~\ref{fig:origishheatmap1} 
and~\ref{fig:origsvmheatmap1}. For comparison, the same process was applied to 
the original datasets to obtain Figures~\ref{fig:ishheatmap1} and~\ref{fig:svmheatmap1}.

As can be seen in Figures~\ref{fig:ishheatmap1} and~\ref{fig:svmheatmap1}, 
the activation regions for spam images from the original ISH and MD datasets 
are skewed towards the top and bottom. The narrow shape of these regions represent 
the regions in spam images that generate the highest activations in 
the neurons of the deep learning-based classifier. 
The central region of the average class activation map for spam images from the 
MD dataset is much darker in comparison to that of spam images 
from the ISH dataset due to the superimposition of natural images 
directly onto spam features, 
as described in~\cite{support_vector_machines_for_image_spam}. 

In contrast, Figures~\ref{fig:origishheatmap1} and~\ref{fig:origsvmheatmap1} indicate 
that the introduction of natural and universal adversarial perturbations 
are able to more evenly distribute the activation regions. This result shows 
that the spam images from the modified datasets are much closer---in terms 
of natural features---to non-spam images. This also suggests that the proposed 
method outperforms the procedure used to generate the original MD dataset 
as outlined in~\cite{support_vector_machines_for_image_spam}. 

\section{Conclusion and Future Work}\label{sec:con}

Modern deep learning-based image spam classifiers can accurately classify 
image spam that has appeared to date in the wild. 
However, spammers are constantly creating new countermeasures to
defeat anti-spam technology. Consequently, the eventual use of adversarial examples 
to combat deep learning-based image spam filters is inevitable. 

In this chapter, four adversarial attacks were selected based on specific restrictions 
and constraints of the image spam problem. These adversarial attacks 
were evaluated on the CNN and MLP architectures introduced in~\cite{convolutional_neural_networks_for_image_spam}. For training data,
we used the dataset presented in~\cite{image_spam_hunter} 
and~\cite{support_vector_machines_for_image_spam}. 
The Fast Gradient Sign Method (FGSM) attack, C\&W's Attack, DeepFool, and the 
Universal Perturbation attack were all evaluated based on mean accuracy reduction, 
mean computation time per adversarial spam image, mean $L_2$ distance from the original spam 
images, and ROC curves of the attacked classifiers. Through further statistical analysis, 
the Universal Perturbation was chosen as a base for our proposed image transformation 
attack, due to its versatility and overall high performance 
in terms of accuracy reduction and computation time.

To maximize the number and intensity of natural features in an attack, the approach 
introduced in~\cite{deepdream} for maximizing activations of certain layers in 
a deep neural network was used. This technique serves to generate sets of ``natural perturbations'' 
from the non-spam subsets of the image spam datasets. These natural perturbations 
were then modified via the class activation maps of all spam images in both datasets. 
The class activations were generated using GradCAM from the two convolutional 
neural networks trained on the ISH and MD datasets. These activation maps allow the regions in spam images recognized to contribute most to the spam classification to benefit from a universal adversarial perturbation. 

Our technique resulted in comparable---if not greater---accuracy reduction as
compared to C\&W's Attack. In addition, our approach is 
computation much more efficient than C\&W's Attack. 
Furthermore, the nature of our attack implies that the only potential 
computational bottleneck is generating the modified natural perturbations.
This aspect of the attack
would not be an issue in practice, unless a spammer generates 
vast numbers (i.e., in the millions) of modified 
adversarial spam images. 

A dataset of modified adversarial 
spam images has been generated by the authors by applying the proposed attack to the spam subsets of the ISH and MD datasets. This dataset will be made freely available to researchers.

Future work will include evaluating the ability of adversarial attack defense 
methods. We will consider defensive distillation against adversarial spam images 
generated with our proposed attack. The goal of this research will be to develop defenses specifically designed for natural perturbation-augmented adversarial spam images. For example,
the subtraction of predicted adversarial perturbations is one path that we intend to pursue.



\bibliographystyle{plain}
\bibliography{references.bib}

\end{document}